\documentclass[aps,prappl,foatfix,reprint,longbibliography,superscriptaddress]{revtex4-1}  % for review and submission
\usepackage{graphicx}  % needed for figures
\usepackage{dcolumn}   % needed for some tables
\usepackage{multirow}
\usepackage{bm}        % for math
\usepackage{amssymb}   % for math
\usepackage{amsmath}   % for math
\usepackage{siunitx}   % for SI units
\usepackage{color}     % for \textcolor 
\usepackage[version=4]{mhchem}
\usepackage{xfrac}

\usepackage{blindtext}

\usepackage{verbatim}
\usepackage[normalem]{ulem} %Strikeout with \sout{} command,
\usepackage{float}

\DeclareSIUnit\torr{Torr}
\DeclareSIUnit\sq{\ensuremath{\Box}}

\begin{document}

% the following line is for submission, including submission to the arXiv!!
%\hspace{5.2in} \mbox{Fermilab-Pub-04/xxx-E}

\title{Intermodulation spectroscopy and the nonlinear response \\ of two-level systems in superconducting coplanar waveguide resonators}

\author{Janka~Biznárová}
\email{jankab@chalmers.se}
\affiliation{Chalmers University of Technology, Microtechnology and Nanoscience, SE-41296, Gothenburg, Sweden}
\author{J.~C.~Rivera~Hern\'{a}ndez}
\affiliation{Department of Applied Physics, KTH Royal Institute of Technology, SE-106 91 Stockholm, Sweden}
\author{Daniel~Forchheimer}
\affiliation{Intermodulation Products AB, Segersta, Sweden}
\author{Jonas~Bylander}
\affiliation{Chalmers University of Technology, Microtechnology and Nanoscience, SE-41296, Gothenburg, Sweden}
\author{David~B.~Haviland}
\affiliation{Department of Applied Physics, KTH Royal Institute of Technology, SE-106 91 Stockholm, Sweden}
\author{Gustav~Andersson}
\email{gandersson@uchicago.edu}
\affiliation{Chalmers University of Technology, Microtechnology and Nanoscience, SE-41296, Gothenburg, Sweden}
\affiliation{Pritzker School of Molecular Engineering, University of Chicago, IL 60637 Chicago, USA}

\vskip 0.25cm

\date{\today}

%% No more than 600 characters
\begin{abstract}

Two-level system (TLS) loss is typically limiting the coherence of superconducting quantum circuits. 
The loss induced by TLS defects is nonlinear, resulting in quality factors with a strong dependence on the circulating microwave power. 
We observe frequency mixing due to this nonlinearity by applying a two-tone drive to a coplanar waveguide resonator and measuring the intermodulation products using a multifrequency lock-in technique. 
This intermodulation spectroscopy method provides an efficient approach to characterizing TLS loss in superconducting circuits.
Using harmonic balance reconstruction, we recover the nonlinear parameters of the device-TLS interaction, which are in good agreement with the standard tunnelling model for TLSs.

\end{abstract}

%\pacs{}
%\keywords{}

\maketitle

\section{Introduction}

Superconducting quantum computing is facing a challenge in the relatively short coherent lifetimes of state-of-the-art qubits, currently in the sub-millisecond range~\cite{Kjaergaard2020, Place2021, Spring2022, Wang2022, Somoroff2023}.  
A significant source of decoherence for these qubits is dielectric loss caused by parasitic two-level system (TLS) defects~\cite{pappas, lisenfeld2015, Muller2019}, whose microscopic origin is a topic of ongoing active research~\cite{molina, Un2022, Spiecker2023}.
These TLSs can couple to qubits, providing a source of energy loss and parameter fluctuations, and therefore decoherence.
In this article we introduce a method for the measurement and analysis of the inherent nonlinearity of these TLSs in superconducting resonators.

Superconducting resonators are essential components of quantum processors and a good testbed for characterizing dielectric loss.
In circuit quantum electrodynamics resonators are treated as linear elements, typically modelled as lumped-element LC oscillators.
The resonator's coupling to TLS defects gives rise to a characteristic dependence of the internal quality factor $Q_i$ on circulating power, with a drop in $Q_i$ at low powers where the TLS bath is not driven to saturation~\cite{sage, Megrant2012}.
This power-dependent response makes resonators ideal for probing TLSs in a qubit's environment.

In this work we explore how the power-dependent behaviour of $Q_i$ translates to a low-power nonlinear response of the resonator. 
We measure this nonlinearity in a high-quality-factor coplanar waveguide (CPW) resonator, showing that it gives rise to frequency mixing, or intermodulation -- a characteristic feature of nonlinear response which has been observed for TLS-induced~\cite{naaman2023} and other nonlinearities in superconducting circuits~\cite{Erickson2014, Dmitriev2019, Weissl2019}. 
Intermodulation results in response at integer combinations of the applied drive frequencies, $f_\mathrm{IMP} = \sum_i k_i f_i$, where $f_i$ are drive frequencies and $k_i$ integers.
The amplitude and phase of these intermodulation products (IMPs) can be measured with large signal-to-noise ratios (SNR) using lock-in techniques.
Modelling the system as a driven harmonic oscillator with nonlinear damping due to parasitic TLSs, we reconstruct the nonlinear response of the device as a function of drive amplitude and frequency using harmonic balance analysis.

\section{Experimental methods}

We perform all measurements on a superconducting aluminium $\lambda/4$ coplanar waveguide resonator with a resonance frequency of \SI{4.11}{\giga\hertz},  capacitively coupled to a transmission line in a notch configuration.
The device has no engineered nonlinear elements.
All microwave measurements presented here were carried out inside a dilution refrigerator at $\sim\!\SI{10}{\milli\kelvin}$ using the set-up outlined in Appendix~\ref{sup:setup}.

\begin{figure*}
    \includegraphics[width=17.5cm]{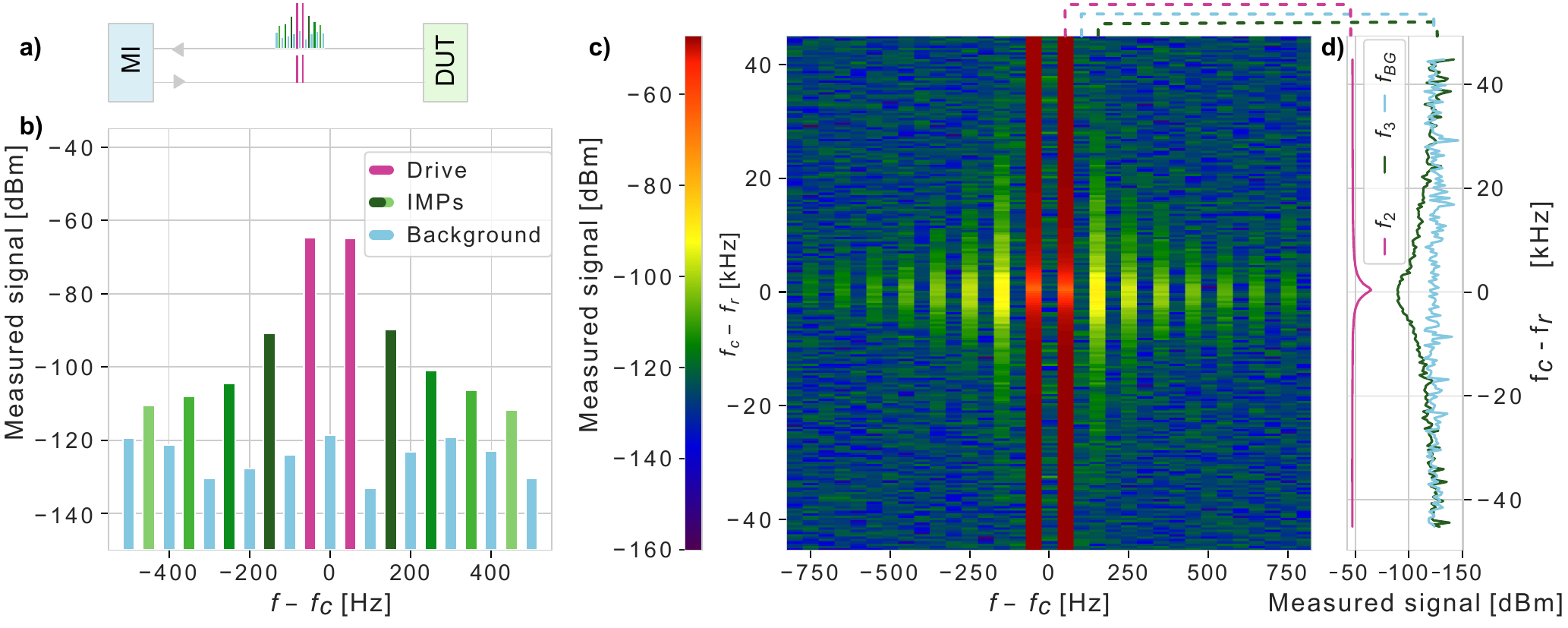}
    \caption{\label{fig:tones} Intermodulation measurements at drive powers corresponding to  $\langle n \rangle = 10^3$ when on resonance. \textbf{a)} A schematic of the tones generated at the output of the measurement instrumentation (MI) and the output of the device-under-test (DUT).
    \textbf{b)} IMPs measured when two drive tones are applied at resonance. The abscissa shows the detection comb frequencies $f$ relative to the centre of the detection comb $f_c = (f_1 + f_2)/2$. IMPs are detected at frequencies fulfilling the $f_\mathrm{IMP}=k_1 f_1+ k_2 f_2$ condition, which falls on every other tone in this detection comb.
    \textbf{c)} IMP power spectroscopy performed by sweeping the frequency comb depicted in a) across a frequency spectrum centered around the resonance frequency $f_r$.
    \textbf{d)} Vertical line-cuts of the IMP spectroscopy data shown in c): the device resonance demonstrates as a dip in the transmission of drive tone $f_{1}$, the third IMP $f_3$ presents as a peak at 2$f_{2}-f_{1}$, and the background noise is shown at a frequency $f_{BG}$ that does not correspond to any IMP. 
    The phase response is shown in Appendix~\ref{sup:more_IMP}.}
\end{figure*}

For the initial characterization of the resonator, we measure the $S_{21}$ scattering parameter in transmission using a vector network analyzer and extract the relevant parameters such as $Q_i$ and the resonance frequency $f_r$ using the circle-fitting method~\cite{probst2015}, as detailed in Appendix~\ref{sup:res_char}.
We characterize $Q_i$ for excitation powers ranging from the single photon level, relevant for quantum computing applications, up to $10^{8}$ photons where we observe other, non-TLS related loss and nonlinear mechanisms.

The average number of photons $\langle n \rangle$ circulating in the resonator is estimated using the relation
\begin{equation}
    \langle n \rangle = \frac{Z_0}{Z_r}\frac{Q_l^2}{|Q_c|}\frac{2 P_{in}}{\hbar(2\pi f_r)^2},
\end{equation}
where $Z_0$ and $Z_r$ are the characteristic impedances of the transmission line and the resonator, respectively. 
The power at the device input $P_{in}$ is calculated by subtracting the input line attenuation from the measurement instrument output power. 
The input line attenuation was obtained via an ac-Stark shift measurement of a qubit device~\cite{bruno2015} in a separate experiment.

To probe nonlinearities in the device through intermodulation distortion we use a multi-frequency lock-in amplifier platform~\cite{Tholen2022}, where we output two drive tones separated by $\Delta = \SI{100}{\hertz}$, see Fig.~\ref{fig:tones}a). 
We choose this drive spacing to be significantly smaller than the power-dependent linewidth of the resonator (8-\SI{11}{\kilo\hertz}). 
This choice guarantees that both the drives and the lowest-order IMPs remain well within the resonance linewidth.

We measure the response of the device to these two drive tones using a detection frequency comb demodulating the IQ quadratures at frequencies spaced by $\Delta/2$. 
Consequently, intermodulation produces a signal at every other frequency in this comb, and the remaining detection frequencies sample the noise background. 
The generation of drive tones and digitization of the response is performed in the second Nyquist zone, without the use of analog mixers for frequency conversion.
Odd-order IMPs, i.e. when $|k_1| + |k_2|$ is an odd number, appear as a comb of frequencies when $f_c$ coincides with the resonance frequency $f_r$, as shown in Fig.~\ref{fig:tones}~b).
Even-order IMPs fall far outside of the resonance linewidth.

We perform intermodulation spectroscopy, stepping the detection frequency comb with the two drives in the centre across a frequency range. 
This frequency range is centred around $f_r$, with a span well outside the resonance linewidth. 
As shown in Fig.~\ref{fig:tones}~c), we detect no IMPs when the centre of the comb is off-resonant.
As the two drives approach $f_r$, we measure a clear signal at the IMP frequencies, peaking at $f_c = f_r$.
The decay of the measured IMPs outside of resonance shows that the IMPs are generated in the device and are not artifacts of nonlinearities present elsewhere in the measurement system.
Thus, we can use this intermodulation spectroscopy method to characterize nonlinearities native to our CPW resonator.

\begin{figure*}
    \centering
    \includegraphics[width=\textwidth]{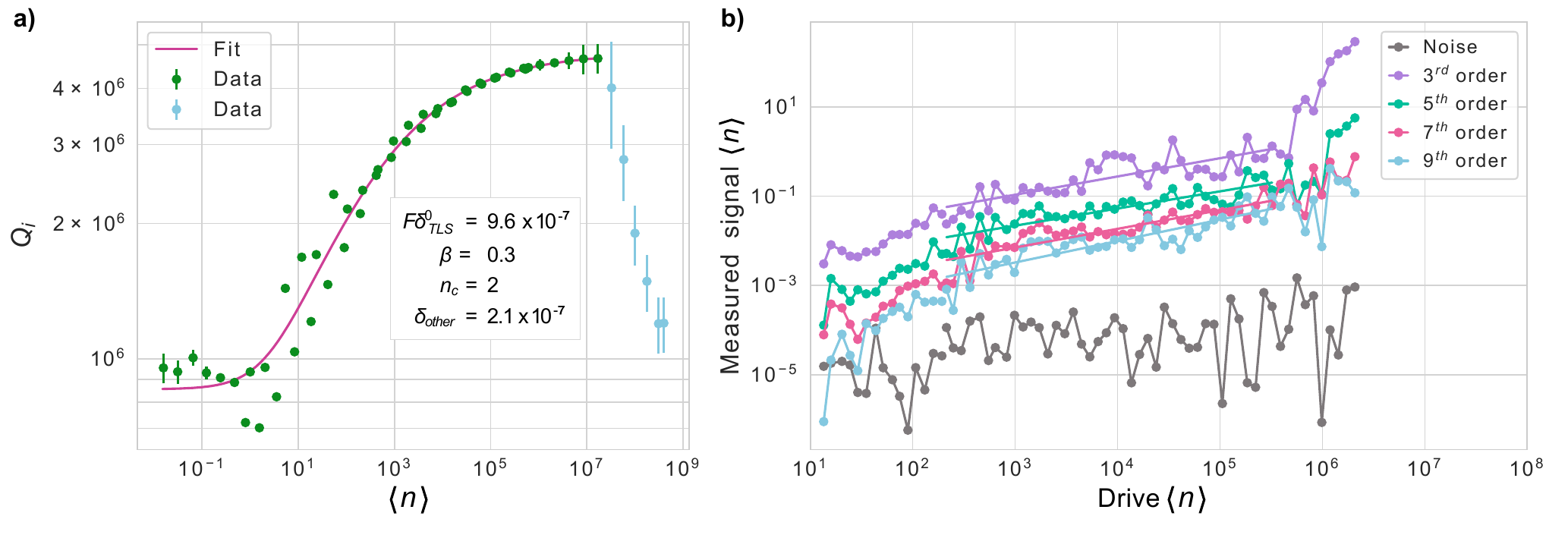}
    \caption{\label{fig:sweeps} 
    \textbf{a)} Internal quality factor $Q_i$ as a function of applied power $\langle n \rangle$, fitted to equation~\ref{eq:tls}. 
    \textbf{b)} The power of the four strongest IMPs, as well as the background noise at non-IMP frequencies, plotted as a function of applied drive power, for the case when the drive tones are applied at resonance.
    In the entirety of the measured power span, the signal corresponding to the two strongest IMPs measured is well above the noise floor.
    At drive powers exceeding $\sim \SI{5e5}{}$ photons, the dominant nonlinearity is due to kinetic inductance, and the strength of the IMPs increases rapidly.
    The slope of the power dependence of the data plotted on a logarithmic scale is fitted for $\langle n \rangle$ in the range of 200--\SI{2e5}{} photons.}
\end{figure*}

\section{Results}

The standard resonator characterization shows the expected trend where $Q_i$ is suppressed at low $\langle n \rangle$ due to interaction with parasitic TLSs. 
This observation fits the so-called standard tunnelling model~\cite{burnett2016}
\begin{equation}\label{eq:tls}
    \frac{1}{Q_i } = F \delta^0_{TLS} \frac{\tanh{(h f_r / 2 k_B T)}}{\left( 1 + \langle n \rangle / n_c\right)^\beta} + \delta_0,
\end{equation}
where $F \delta^0_{TLS}$ describes the power-dependent loss due to parasitic TLSs, with $F$ quantifying the TLS filling factor, and $\delta^0_{TLS}$ is the dielectric loss tangent, dependent on the density of the TLSs. 
The parameter $n_c$ is the critical photon number for the saturation of the TLS bath, and the power-independent term $\delta_0$ quantifies the other, non-TLS-related sources of loss, such as radiation or resistive losses due to quasiparticles. 
$T$, $h$, and $k_B$ denote temperature, the Planck constant, and the Boltzmann constant, respectively.

In the standard tunnelling model, the parameter $\beta = 0.5$ characterizes the  change of $Q_i$ with photon number. 
This value often fits data poorly, especially for high-quality resonators, which show a weaker power-dependence~\cite{burnett2017, niepce2019, marina2020}.
We treat $\beta$ as a fitting parameter, finding a best-fit value $\beta=0.3$ (see Fig.~\ref{fig:sweeps}a). 
These deviations in $\beta$ have been attributed to the spectral instability of the device-TLS system, and the interactions between different TLSs themselves, not just interactions between a TLS and the quantum circuit~\cite{burnett2014, faoro2015, Muellertls}.

Having performed standard resonator characterization measurements, we proceed to intermodulation spectroscopy as a function of applied drive power. 
Figure~\ref{fig:sweeps}b) shows a clear signal at IMP frequencies across the entire measured power range, with a power-dependent amplitude of the measured IMPs. 
We identify three separate regions with different power-dependent characteristics, consistent with the observations from Fig.~\ref{fig:sweeps}a).

The drive powers used in this measurement exceed the critical photon number $n_c$.
In the low-power regime where $\langle n \rangle < 200$, TLSs dominate the total loss of the resonator, and the dependence of the IMPs on power is strong with a variable slope.
In the intermediate drive power regime where $\langle n \rangle \gg n_c$,  the four strongest IMPs show a power-law behaviour, scaling with drive power as $\langle n\rangle^k $.
The fit parameters $k$ are laid out in Table~\ref{tab:pow_fit}, alongside the parameters predicted by the model in section IV. 
We note that due to the saturation of the TLS bath, the IMPs increase more slowly than the drive amplitude (i.e. $k < 1$). 
The opposite is typically the case for intermodulation distortion in electronic devices~\cite{Zumbahlen2008}. 
At drive powers exceeding $\sim$ \SI{5e5}{} photons, the nonlinearity increases sharply, which relates to the onset of kinetic inductance effects at high input power~\cite{Mattis1958, Zmuidzinas2012}. 

These trends are consistent with the $Q_i$ vs. $\langle n \rangle$ curve shown in Fig.~\ref{fig:sweeps}a).
Here, TLSs contribute significantly to the loss at low excitation numbers and become saturated with stronger probes. 
At high power, current-induced pair breaking changes the kinetic inductance and the frequency of the resonator shifts, as shown in Fig.~\ref{fig:fr_vs_n}.

\begin{figure}
    \includegraphics[width=.5\textwidth]{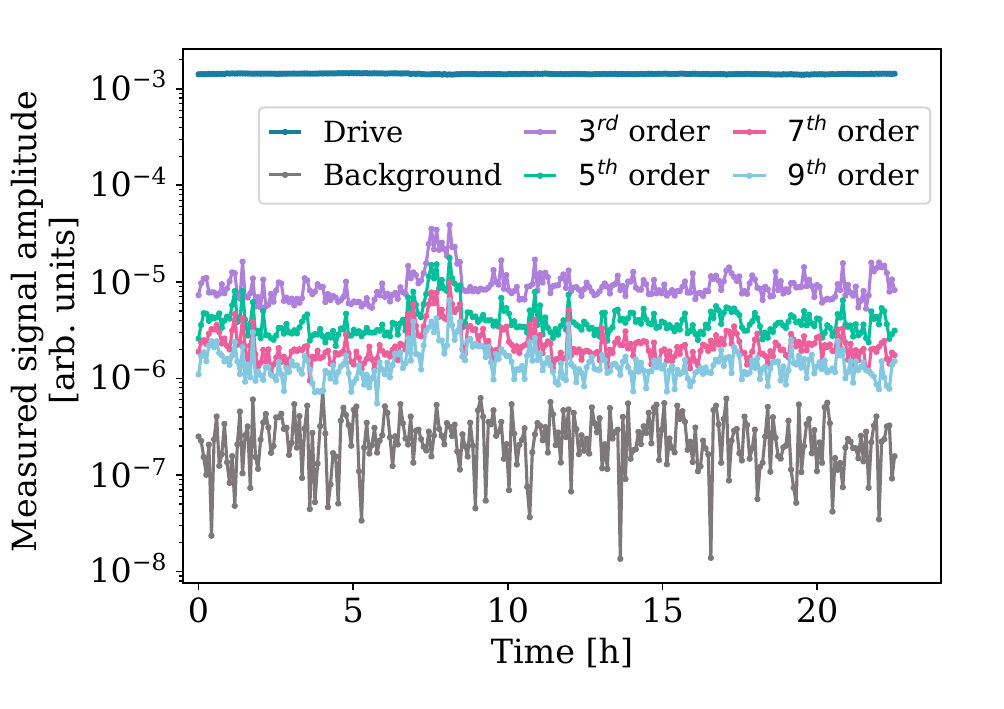}   
    \caption{\label{fig:time} IMPs measured repeatedly for 25h, in 5-minute intervals, at a constant drive power level of $\sim 10^3$ photons. }
\end{figure}

The coupling between quantum circuits and TLSs tends to fluctuate on slow time scales \cite{Burnett2019, Muellertls}.
We study this effect on the measured IMPs in Fig.~\ref{fig:time}, where we repeat the same IMP spectroscopy measurement at a fixed drive power level of $\sim 10^3$ photons for 25 hours. 
Each of the individual measurements is averaged for $\sim$\SI{5}{\minute}. 
We do observe seemingly random fluctuations of the measured IMPs over time in this measurement, correlated between the different orders of IMPs.
The fluctuations do not extend to the measured amplitudes of the drive tones, which are stable in comparison, nor to the measured background noise. 
From this we conclude that the fluctuations are native to the nonlinearity in the device, and do not originate in the measurement setup from sources such as drifts in room-temperature electronics.

We also conclude that the fluctuations shown in Fig.~\ref{fig:sweeps} are likely not a function of applied power.
Rather, the intermodulation spectrum is unstable, revealing a change of coupling between the TLS and the device over time, a typical feature of TLS interaction~\cite{Niepce2021}.
%Instead, they occur over time - a typical feature of the spectrally unstable TLS-interaction behaviour~\cite{Niepce2021}. 
%The fluctuations in IMP amplitude reveal the coupling behaviours between TLS defects and superconducting devices over time.

\begin{table}
    \centering
    \caption{Power dependence of the measured IMPs in Fig.~\ref{fig:sweeps}d), and of the modeled IMPs in Fig.~\ref{fig:sim-power},  fitted to $\langle n\rangle^k \cdot 10^{l}$.}
    \label{tab:pow_fit}
    \begin{tabular}{c|c|c}
        IMP order & Experimental fit  $k$ & Model fit  $k$ \\ \hline
         3 & 0.41 $\pm$ 0.05 & 0.42 \\
         5 & 0.38 $\pm$ 0.04 & 0.43 \\
         7 & 0.42 $\pm$ 0.04 & 0.44 \\
         9 & 0.48 $\pm$ 0.05 & 0.45
    \end{tabular}
\end{table}

\section{Nonlinear parameter reconstruction}

In this section, we detail the modelling and reconstruction of the nonlinear loss due to parasitic TLSs.
Our reconstruction algorithm is based on fitting the linear and nonlinear oscillator parameters to a single intermodulation spectrum by harmonic balance~\cite{Yasuda1988, Hutter2010}.

We model our resonator as a driven harmonic oscillator with non-linear damping caused by parasitic TLSs.
The Heisenberg-Langevin equation of motion describing the dynamics of the resonator in the rotating-wave approximation is then
\begin{equation}
    \dot{\hat{a}} = -i \omega_0 \hat{a} - \frac{\kappa_0 + \kappa_{ext}}{2} \hat{a} - \dfrac{\kappa_{TLS}(|a|)}{2} \hat{a} - \sqrt{\kappa_{ext}} a_{in},
    \label{eq:TLS-model}
\end{equation}
where $\hat{a}$ and $a_{in}$ are the time-dependent intra-cavity modes and the input drives respectively, $\omega_0$ the resonant frequency, $\kappa_0$ the internal linear loss rate, $\kappa_{ext}$ the external coupling to the transmission line, and $\kappa_{TLS}(|a|)$ the power-dependent loss due to parasitic TLSs.
We consider a power-dependent damping of the standard form of Eq.~\eqref{eq:tls}, giving
\begin{equation}
    \kappa_{TLS}(|a|) a = \dfrac{\kappa_{TLS}}{\left[ 1 + \left( {|a|}/{a_c} \right)^2 \right]^\beta},
    \label{eq:non-linear_damping}
\end{equation}
where $\kappa_{TLS}=f_r F\delta^0_{TLS}$ represents the TLS loss rate, and $a_c^2 = n_c$ the critical photon number.

We drive the system with two tones, separated by $\Delta$, such that the IMPs created are within the resonator's linewidth, where the sensitivity is enhanced.
When the drive frequencies are periodic in some measurement time $T = 2\pi/\Delta$, and the nonlinearities and drives are sufficiently weak, the system's response is a discrete spectrum at frequencies that correspond to integer multiples of $\Delta$.

\begin{figure}
    \includegraphics[width=.5\textwidth]{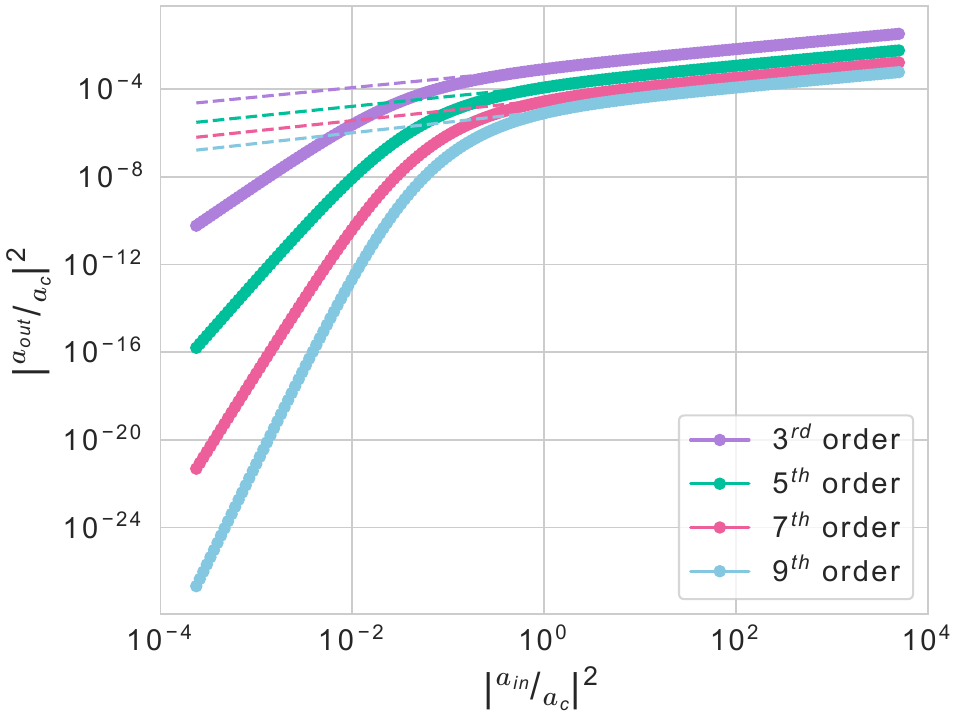} 
    \caption{\label{fig:sim-power} Simulation of the power of the generated IMPs as a function of applied drive power. 
    The parameters of the fit shown with dashed lines are shown in Table~\ref{tab:pow_fit}.}
\end{figure}

We first solve the forward problem by numerically integrating Eq.~\eqref{eq:TLS-model} using scipy.integrate.ode from the SciPy library~\cite{2020SciPy-NMeth} to obtain the time-dependent intra-cavity field.
The output field is then obtained using the input-output relation $\hat{a}_{out} = \sqrt{\kappa_{ext}} \hat{a} + a_{in}$.
Fig.~\ref{fig:sim-power}) shows the four strongest IMPs as a function of drive power simulated with experimentally relevant parameters. 
For strong drives $\langle n \rangle \gg 1$, the IMP power shows a power-law scaling proportional to $\langle n \rangle^k $, with parameters $k$ laid out in Table~\ref{tab:pow_fit}.
This behaviour matches the experimental results shown in Fig.~\ref{fig:sweeps}~b).
We also simulate a sweep of the drive tone frequency across the resonance and analyze the power dependence of the IMPs as a function of $\beta$ (see Appendix~\ref{sec:timesweep} for details).

To reconstruct the nonlinear parameters of the system, we expand the TLS damping in Eq.~\ref{eq:non-linear_damping} in a power series, transforming the equation of motion to
\begin{equation}
    \dot{\hat{a}} = -i \omega_0 \hat{a} - \sum_{n=1}^N c_n |a|^{n-1} \hat{a} - \sqrt{\kappa_{ext}} a_{in},
    \label{eq:Poly-model}
\end{equation}
which is linear in the coefficients $c_n$.
The coefficients are independent of the proposed model, allowing the application of the reconstruction algorithm to various types of nonlinearities~\cite{Forchheimer2012, Weissl2019}, including unknown ones.

In the experiment we measure only a limited number of frequencies with good SNR, corresponding to the two drive tones and 14 IMPs.
These frequencies, denoted as $\omega_k$ with Fourier component $\hat{A}_k$, represent a partial spectral response of the system.
Specifically, $\hat{A}_k$ corresponds to the intracavity field amplitude at the frequency $\omega_k$.

By Fourier transforming and rewriting Eq.~\eqref{eq:Poly-model} in matrix form, we arrive at an expression linear in the coefficients $c_n$~\cite{Weissl2019}
\begin{equation}
    \sum_l H_{kl} p_l = -A_{in, k},
    \label{eq:EOM-matrix}
\end{equation}
where the $k$th row of the matrix $H_{kl}$ is given by
\begin{equation*}
    H_k = 
    \begin{pmatrix}
        i \omega_k \hat{A}_k & i \hat{A}_k & \hat{A}_k & \mathcal{F} \left\{ |a|^{2} \hat{a} \right\}_k & \cdots & \mathcal{F} \left\{ |a|^{N-1} \hat{a} \right\}_k
    \end{pmatrix},
\end{equation*}
and $p_l$ are the components of a vector containing the unknown parameters
\begin{equation*}
    \sqrt{\kappa_{ext}} \; p^T = 
    \begin{pmatrix}
        1 & \omega_0 & c_1 & \cdots & c_N    
    \end{pmatrix}.
\end{equation*}

We solve Eq.~\eqref{eq:EOM-matrix} by numerical pseudo-inverse of the matrix $H$ to find the best-fit vector $p$, which includes the polynomial coefficients $c_n$.
To extract the TLS damping parameters $\kappa_{TLS}$, $a_c$ and $\beta$ from the polynomial expansion coefficients and to compare our harmonic balance results with the circle fit analysis, we perform a least-squares fitting.
For this fit, we set $\kappa_0=0.86$~kHz, based on the value extracted from a fit of the parameters extracted from the standard circle fit to Eq.~\ref{eq:tls} in the high-power regime.
Our reconstruction is based on measurements of a single intermodulation spectrum and encounters limitations in simultaneously capturing both $a_c$ and $\beta$.
We therefore fix $\beta=0.3$ motivated by the slope of the 3rd-order IMP strength in the TLS saturation regime (see Appendix~\ref{sec:timesweep}).

The resulting parameters, listed in Table~\ref{tab:rec_values}, show good agreement with the standard fit values. 
Without fixing the parameter $\beta$, our reconstruction method still yields the important TLS loss parameter $\kappa_{TLS}$ in good agreement with the standard method.
Differences may be due to variable conditions across cooling cycles as the standard resonance sweeps and IMP data were obtained in separate cooldowns~\cite{McRae2021}.

\begin{table}
    \centering
    \caption{Reconstructed resonator parameters via fitting the standard tunnelling model Eq.~\eqref{eq:tls} and the intermodulation spectral method with harmonic balance.}
    \begin{tabular}{c|c|c}
         Parameters & Standard model & Harmonic balance \\ \hline
         $f_0$ [GHz] & 4.11 & 4.11 \\
         %$\kappa_0$ [kHz] & 0.86 & 0.86 \\
         $\kappa_{ext}$ [kHz] & 6.77 & 6.62 \\
         $\kappa_{TLS}$ [kHz] & 3.95 & 5.22 \\
         $a_c$ [$\sqrt{n}$] & 1.414 & 1.641 \\
    \end{tabular}
    \label{tab:rec_values}
\end{table}

\section{Discussion and conclusion}

We demonstrate the inherent nonlinear nature of parasitic TLSs in superconducting circuits by observing frequency mixing products in a CPW resonator when driven near-resonantly by two slightly detuned drive tones. 
We characterize the intermodulation products of this frequency mixing as a function of drive detuning with respect to the resonance frequency, as well as the applied drive power.
Drawing a comparison to the standard tunnelling model for TLSs, we observe an analogous power dependence of the IMPs on applied drive power, as we expect from the conventional analysis of the $Q_i$ suppression at decreasing drive powers as an effect of a saturable bath of TLSs.

Repeated measurements of the IMPs generated in the device when driven by the same conditions over \SI{25}{\hour} show correlated jumps in the IMP amplitudes.
This approach provides an efficient method of studying the timescales of TLS-device interaction with a high SNR.

Using harmonic balance analysis and assuming a model for a driven harmonic oscillator with nonlinear damping, we reconstruct the nonlinear parameters of this device-TLS interaction, in good agreement with the parameters extracted from the experimental data using the conventional tunnelling TLS model. 
Our reconstruction method is promising for characterizing TLSs in superconducting devices, as it eliminates the need for a comprehensive frequency and power sweep, enabling analysis through a single intermodulation spectrum.

\begin{acknowledgments}

This work was funded by the Knut and Alice Wallenberg (KAW) Foundation through the Wallenberg Center for Quantum Technology (WACQT). 
The device fabrication was performed at Myfab Chalmers.
D. B. H. and D. F. are part owners of the company Intermodulation Products AB, which produces the digital microwave platform used in this experiment.

\end{acknowledgments}

\appendix

\section{Sample fabrication} \label{sec:appendix_fab}

The device was fabricated on a \SI{280}{\micro\meter} thick, high resistivity ($\rho \geq \SI{10}{\kilo\ohm\centi\meter}$), intrinsic silicon substrate. 
Prior to thin film deposition, the substrate was submerged in 2\% HF for $\SI{60}{\second}$ to remove the native surface oxide, then rinsed in deionized water and loaded immediately into a heated load-lock of a sputter deposition tool, where the substrate was heated to 80°C during pump-down. 
After transfer into the deposition chamber, the substrate was heated to \SI{300}{\celsius} for 10 minutes and left to cool down for 16 hours, after which the chamber pressure was \SI{2.2e-8}{\milli \bar}.
A \SI{150}{\nano\meter} thick layer of Al was deposited by sputtering at \SI{0.9}{\nano\meter\per\second}, and the film was oxidized in situ by static oxidation at \SI{10}{\milli\bar}. 

The device was patterned by direct-write optical lithography on a PMMA A2 + S1805 resist stack. 
The optically sensitive S1805 was developed in a TMAH-based developer.
After development, the exposed PMMA protecting the underlying Al from TMAH was ashed in oxygen plasma, and the pattern was transferred into Al by wet etch in an \ce{H3PO4}:\ce{HNO3}:\ce{CH2O2} solution at room temperature.

\begin{figure*}[t]
    \includegraphics[width=.8\textwidth]{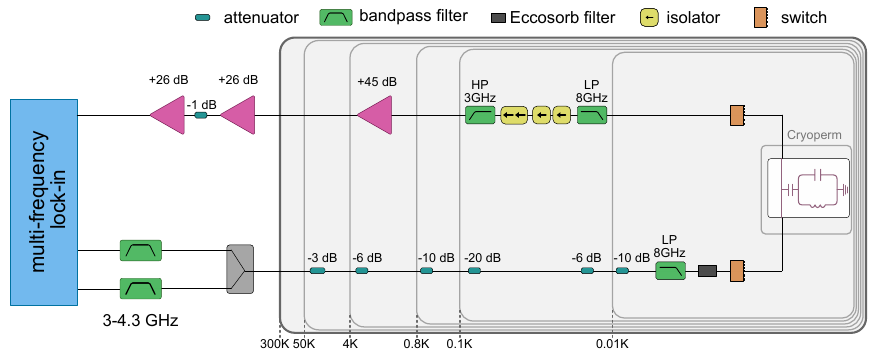}
    \caption{\label{fig:sup_setup} Cryogenic measurement setup. }
\end{figure*}

The remaining resist was then dissolved in an NMP-based resist remover, acetone, and isopropanol. 
The wafer was coated with a fresh layer of protective photoresist (S1805), diced, and stripped again in the aforementioned solvents. The cleaning was finished with an ozone-cleaning step.

\section{\label{sup:setup}Measurement setup}

The chip is placed in a light-tight copper sample box, where the transmission line is wire-bonded to input and output SMA connectors, and the ground plane of the CPW is wire-bonded to the copper box along the chip periphery.

The sample box is then mounted onto a copper tail attached to the mixing chamber stage of a dilution refrigerator. 
The copper tail is enclosed in an additional copper can coated with a layer of Stycast mixed with SiC on the inside, which is in turn enclosed in a Cryoperm magnetic shield. 
This is in addition to the shields attached to every temperature stage as marked in Fig.~\ref{fig:sup_setup}.

On the input signal line, the incoming signal is attenuated at each stage with individual attenuators amounting to a total of  \SI{55}{\decibel} attenuation. 
The input line attenuated by an additional \SI{11}{\decibel}, as determined from an ac-Stark shift measurement of a qubit.

Microwave switches are present below the mixing chamber stage on both the input and output lines. 
On the output line, above the \SI{10}{\milli\kelvin} stage, the signal is passed through a chain of microwave isolators and filters to suppress reflected signals, as well as thermal noise coming from the higher temperature stages and the amplifiers.
The weak signal is amplified at the \SI{4}{\kelvin} stage by a low-noise HEMT cryogenic amplifier.
At room temperature, the signal is further amplified by two gain block amplifiers connected in series.

For intermodulation product measurements, the cryogenic microwave measurement setup was connected to a multi-frequency lock-in amplifier~\cite{Tholen2022}.
We use a separate signal output for each drive tone, passing through a 3-4.3 GHz bandpass filter before being combined in a power splitter and routed through the cryogenic measurement setup.  

For the quality factor characterization of the device, the microwave setup was connected to a vector network analyzer, with a \SI{30}{\decibel} attenuator added to the output of the analyzer at low power measurements.

\section{\label{sup:more_IMP}Additional data}

To investigate the origins of the non-linearity we observe through intermodulation, we repeat the intermodulation measurement; only this time, we place the frequency comb far outside the resonance, detuned by \SI{700}{\kilo\hertz}. 
Here the resonator does not interact with the drive signals, and instead, we measure the transmission of our microwave measurement set-up and the transmission line of the device itself. 
We do not observe any intermodulation here, ruling out the measurement set-up as the origin of this nonlinearity.

\begin{figure}
    \centering
    \includegraphics[width=.45\textwidth]{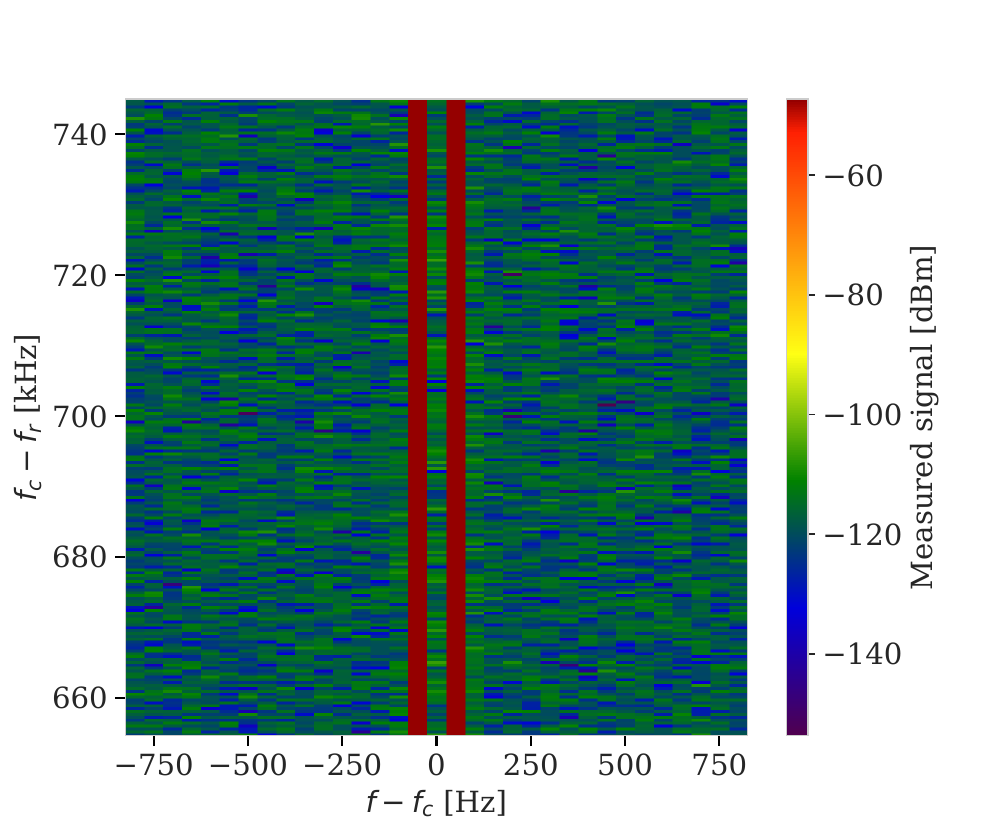}
    \caption{\label{fig:CM_detuned}
    Intermodulation measurement performed around a frequency \SI{700}{\kilo\hertz} away from resonance. 
    We observe no intermodulation products as a result of the two drive tones, showing that the non-linearities we observe are native to the device and not an artifact of our measurement set-up.}
\end{figure}

\begin{figure}
    \centering
    \includegraphics[width=.45\textwidth]{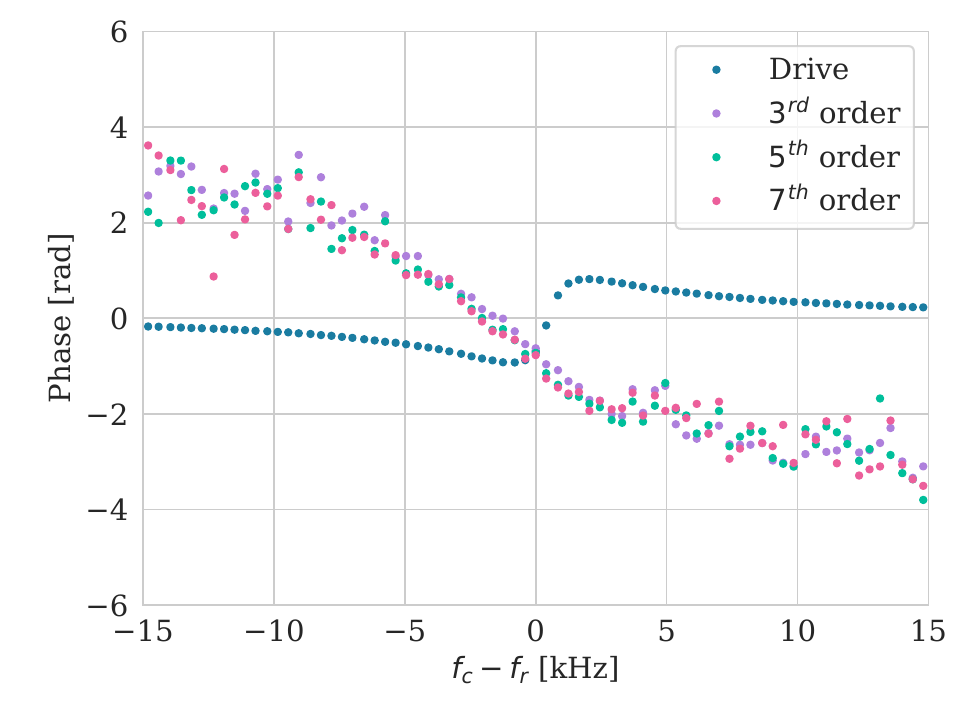}
    \caption{\label{fig:Phase}
    The phase response of the resonator at the drive frequency and the three strongest IMPs, as a function of drive frequency. 
    The drives are at resonance when the detection comb frequency $f_c$ coincides with the resonance frequency $f_r$.}
\end{figure}

We show the phase response of the resonator and the three strongest IMPs in Fig.~\ref{fig:Phase}. 
The phase response at the drive tone wraps by less than $\pi$ across resonance, which is an expected response for an asymmetric, overcoupled  resonator in a notch configuration~\cite{simoen_phd}. 
The phase response of the IMPs wraps by $2\pi$.

\section{\label{sup:res_char}Resonator characterization}

\begin{figure}
    \centering
    \includegraphics[width=.47\textwidth]{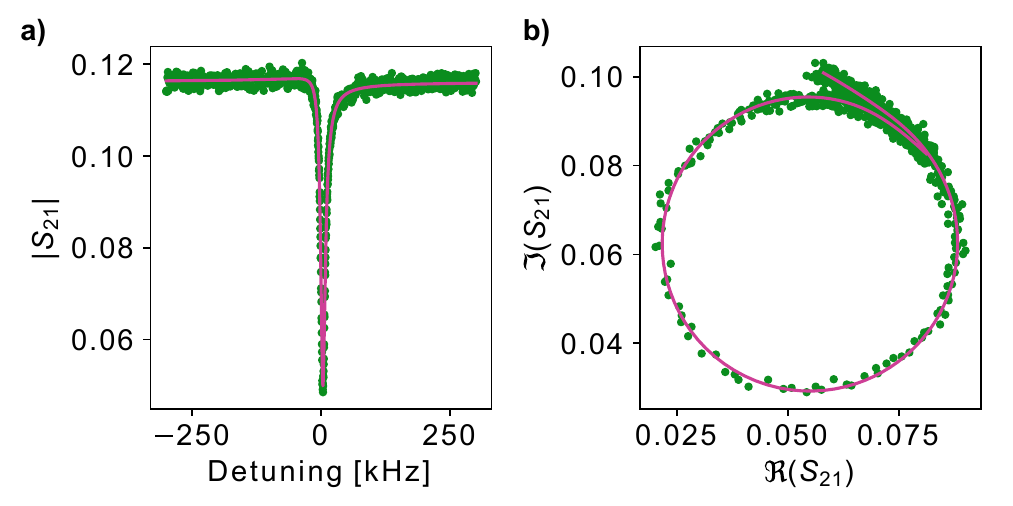}
    \caption{\label{fig:circlefit}
    The resonance circle fit method at the single photon level. 
    \textbf{a)} shows the fit of the absolute value of the transmission scattering parameter $S_{21}$ as a function of frequency, with a dip in transmission at resonance.
    \textbf{b)} shows the resonance circle on the complex plane.}
\end{figure}

The initial characterization of the resonator is performed using a vector network analyzer, where the complex $S_{21}$ transmission scattering parameter is measured in a spectroscopy. 
The $S_{21}$ data is analyzed using the resonance circle fitting method with diameter correction~\cite{probst2015}:
\begin{equation}
    S_{21} (f) = ae^{i\alpha}e^{-2\pi f \tau} \left[1 - \frac{(Q_l/|Q_c|) e^{i\phi}}{1 + 2iQ_l (f/f_r-1)} \right]
\end{equation}
where $f_r$ is the resonance frequency, $f$ the probe frequency, $\phi$ the impedance mismatch, $Q_c$ the coupling quality factor and $Q_l$ the loaded quality factor, from which $Q_i$ can be extracted with $Q_l^{-1} = Q_i^{-1} + Re\{Q_c^{-1}\}$.
The remaining parameters, additional amplitude $a$, phase shift $\alpha$, and electronic delay $\tau$, account for environmental imperfections.
For an example of the accuracy of this fit at the single photon level, see Fig.~\ref{fig:circlefit}.

The circle fitting method is at its most reliable when the resonator is near-critically coupled to the transmission line, i.e. $Q_i \approx |Q_c|$.
With a $|Q_c|$ of $\SI{6e5}{}$ and a $Q_i$ in the range of 1--\SI{5e6}{} depending on their applied drive power, our device is close to this condition, as seen in the symmetry of the resonance circle on the complex plane plotted in Fig.~\ref{fig:circlefit}~b).

\begin{figure}
    \centering
    \includegraphics[width=.45\textwidth]{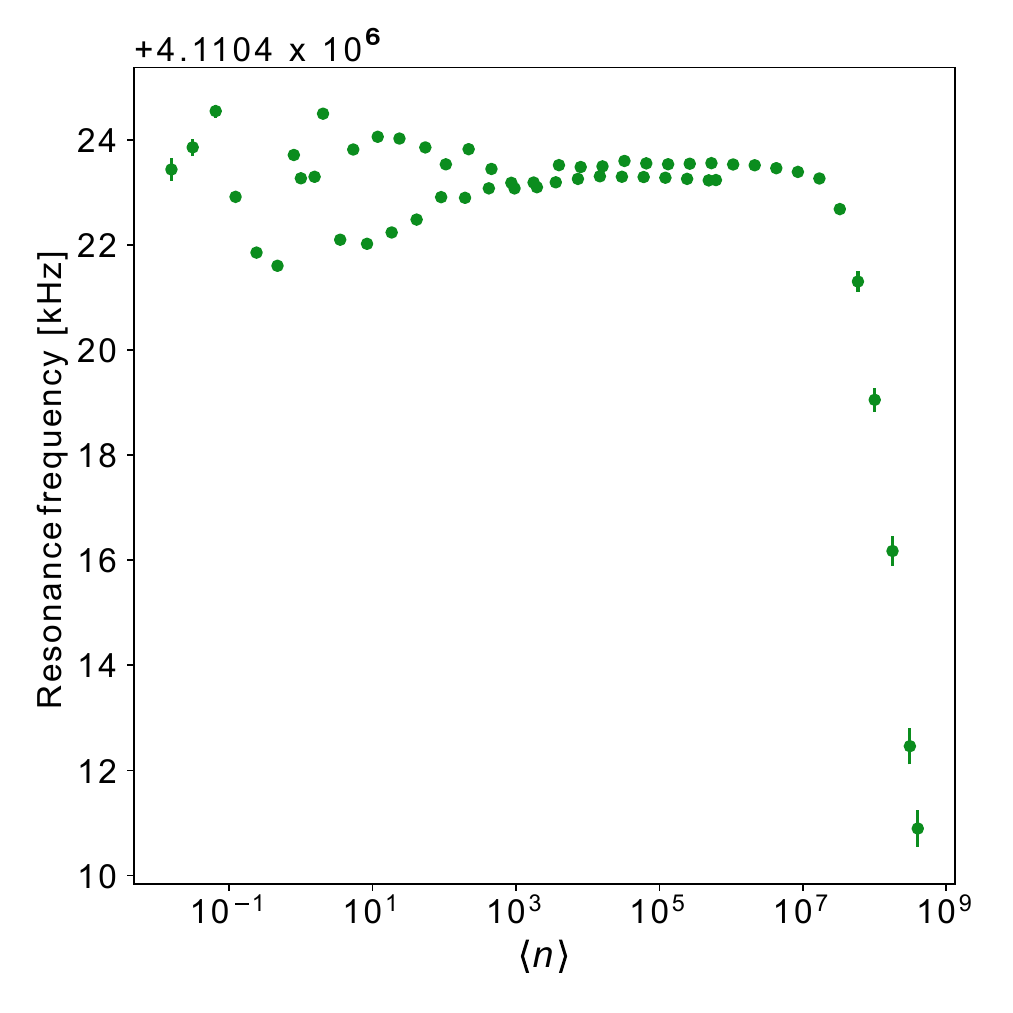}
    \caption{\label{fig:fr_vs_n}
    The resonator's frequency extracted from a circle fit of the complex $S_{21}$ data, as a function of average photon number $\langle n \rangle$}
\end{figure}

\section{Additional modeling} \label{sec:timesweep}

\begin{figure}[!t]
    \centering
    \includegraphics[width=\textwidth]{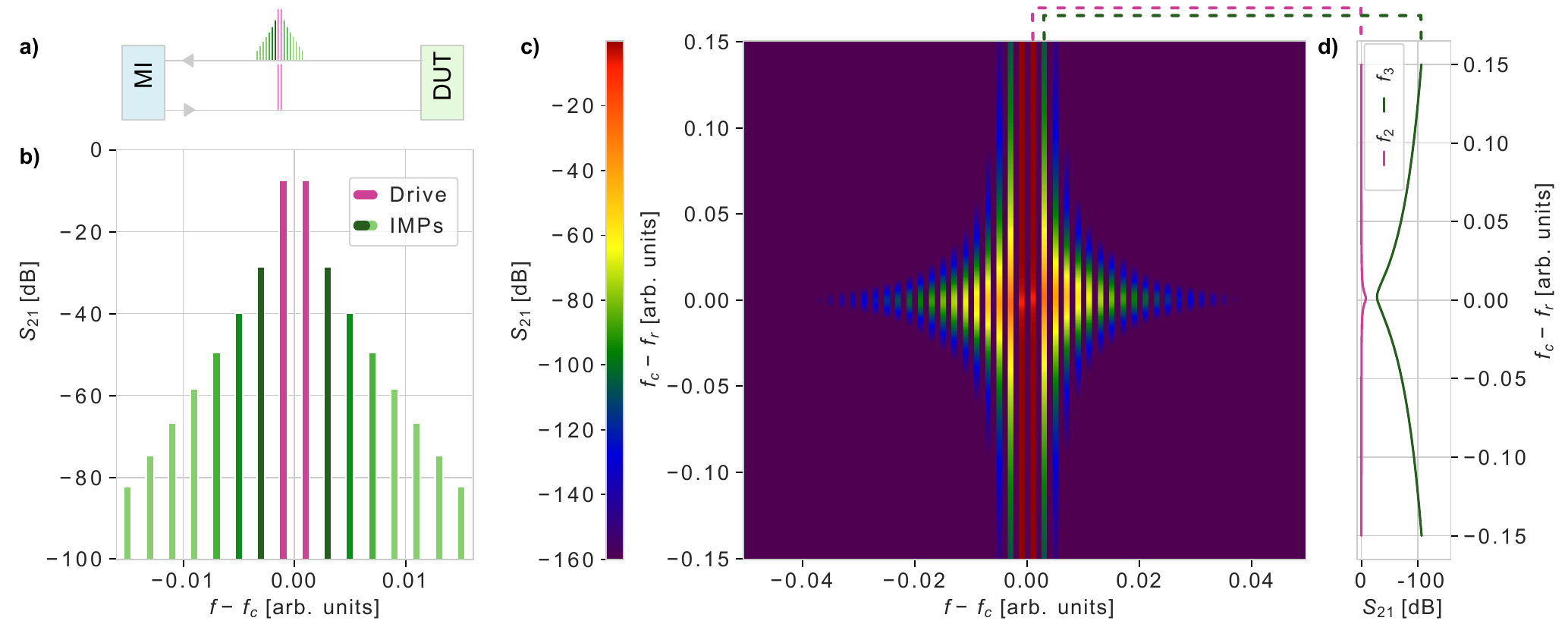}
    \caption{\label{fig:tones_sim} Intermodulation simulation results. \textbf{a)} A schematic of the input tones and the output of the device-under-test (DUT).
    \textbf{b)} Simulated IMPs when two drive tones are applied inside the resonance linewidth. The abscissa shows the detection comb frequencies $f$ relative to the centre of the detection comb $f_c = (f_1 + f_2)/2$. IMPs appear at frequencies fulfilling the $f_\mathrm{IMP}=k_1 f_1+ k_2 f_2$ condition, which falls on every other tone in this detection comb.
    \textbf{c)} Simulated IMP spectroscopy performed by sweeping the frequency comb depicted in a) across a frequency spectrum centred around the resonance frequency $f_r$. 
    \textbf{d)} Vertical line cuts of the simulated IMP spectroscopy data shown in c). The device resonance shows a dip in the transmission of drive tone $f_{1}$, and the third IMP $f_3$ presents a peak at 2$f_{2}-f_{1}$.
    }
\end{figure}

\begin{figure}[!htb]
    \centering
    \includegraphics[width=.4\textwidth]{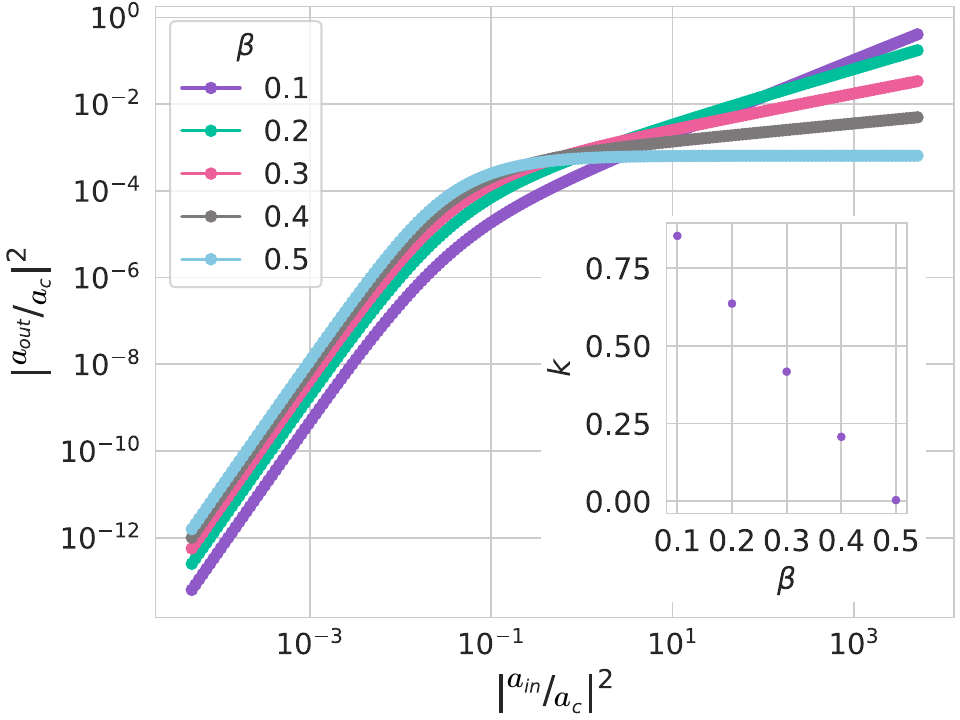}
    \caption{\label{fig:Sim-beta} Simulation of the power of the 3rd order IMP as a function of the input drive power normalized by photon number at different $\beta$ values.
    The inset shows the value of the slope of the TLS saturation regime $k$ as a function of $\beta$.
    }
\end{figure}

We simulate a sweep of the drive frequencies across the resonance shown in Fig.~\ref{fig:tones_sim}.
The drive strength presents a dip on resonance, while the IMP strength increases, qualitatively matching the experimental results in Fig.~\ref{fig:tones}.
We also simulate the power dependence of the third-order IMP for various $\beta$ values, as illustrated in Fig.~\ref{fig:Sim-beta}.
Notably, with increasing $\beta$ we observe a reduction in the slope of the third-order IMP above $n_c$.
The connection between the slope above $n_c$ and the parameter $\beta$ is seen in the inset of Fig.~\ref{fig:Sim-beta}.
As $\beta$ increases, we observe a noticeable decrease in the slope, becoming 0 at  $\beta=0.5$.
This behaviour indicates that the observed power dependence above $n_c$ is not a feature of the standard tunnelling model with $\beta=0.5$.
The value of  $\beta=0.3$ found in Fig.~\ref{fig:sweeps} is also in agreement with this simulation.

\onecolumngrid

\twocolumngrid

%\vfill

%\bibliographystyle{apsrev4-1}
%\bibliographystyle{unsrt}
\clearpage

\bibliography{refs.bib}

\end{document}